\documentclass{aa}
\usepackage{times}
\usepackage{graphics}
\begin{document}

\thesaurus{2(02.03.1), 03.13.4, 05.03.1 }
\title{The power spectrum of geodesic divergences \\
       as an early detector of chaotic motion}
\author{Ch.L. Vozikis, H. Varvoglis and K. Tsiganis}
\offprints{Ch.L. Vozikis}
\mail{chriss@astro.auth.gr}
\institute{Aristotle University of Thessaloniki, Department of
Physics, Section of Astrophysics, Astronomy and Mechanics,
\\ 54006 Thessaloniki, Greece}
\date{Received ......, 1999 / Accepted ......, 2000} 

\titlerunning{PSOD -- a new chaos detector}
\authorrunning{Vozikis et al.}
\maketitle

\begin{abstract}
We propose a new method for determining the stochastic
or ordered nature of trajectories in non-integrable Hamiltonian dynamical systems.
The method consists of constructing a time-series from the divergence
of nearby trajectories 
and then performing a power spectrum analysis of the series. Ordered
trajectories 
present a  spectrum that consists of a few spikes
while the spectrum of stochastic trajectories is continuous. A test of
the method with three 
different systems, a 2-D mapping as well as a 2-D and a 3-D Hamiltonian,
shows that the method is fast and efficient, even in the case
of sticky trajectories.
The method is also applied to the motion of asteroids in the Solar
System.

\end{abstract}


\section{Introduction}

The problem of distinguishing a chaotic from an ordered trajectory in
a non-integrable Hamiltonian system has been a topic of active
investigation since the pioneering work of H\'{e}non and Heiles (1964). 
Initially, when the study was restricted to 2-D
systems, the work was done through
surface of section plots. Later, when systems with more than 
2-D were considered, the method of choice was the calculation of the
Lyapunov Characteristic Numbers (LCNs) (Benettin et al. 1976,
Froeschl\'e 1984). Unfortunately both the above methods suffer from
the same drawback, namely they are not able to distinguish easily an 
ordered from a ``sticky'' chaotic trajectory.  Various methods have been
devised since then to address the above problem, namely the
distinction of an ordered from a chaotic trajectory using a relatively
short-time trajectory segment.

These methods generally fall into two main classes: those that use
frequency or correlation analysis of a time-series, constructed by the
values of generalized co-ordinates (or
functions of them), and those that
use the geodesic
divergence of initially nearby trajectories. In the first
class belong the ``old'' method of the rotation
number (Contopoulos 1966), 
the frequency map analysis developed by Laskar  (Laskar et al. 1992,
Laskar 1993) and the power spectrum analysis of
quasi-integrals developed by
Voyatzis \& Ichtiaroglou (1992). 
In the second belong the probability density function analysis of
stretching numbers
developed by Froeschl\'e et al. (1993) and Voglis \& Contopoulos
(1994) and 
the Fast Lyapunov Indicators method developed by
Froeschl\'e et al. (1997). Each one of the above methods has its own
advantages and weaknesses; in particular some of them are more
suitable to test {\it large sets of trajectories} 
rather than {\it single ones}, some are more efficient for {\it 2-D
systems} rather than for {\it N($>$2)-D systems} 
and some perform better for {\it mappings} rather than {\it flows}.

In 1997,  Contopoulos
\& Voglis introduced a new method for
distinguishing chaotic from 
ordered trajectories, which does not belong to any of the above
mentioned two classes but, 
instead, may be classified as ``mixed''. This new method is based on the 
analysis of the probability 
density of {\it helicity} and {\it twist angles}. Voglis \&
Efthymiopoulos (1998) and subsequently Froeschl\'e \& Lega (1998) showed 
that the twist angles method is very efficient in testing {\it phase space
regions}, at least in cases of 
2-D systems, where the twist angles can be easily calculated.
More recently Voglis et al. (1998, 1999) proposed two new and very
efficient methods, namely the method of ``Dynamical Spectral
Distance'' (DSD), which is particularly suitable for the characterization 
of single trajectories in 4-D maps, and the method of ``Rotational Tori 
Recognizer'' (ROTOR) which is very efficient for testing wide areas of the 
phase space of 2-D maps.

In the present paper we are introducing a new ``mixed" method, which we
show that it is at least as 
sensitive as the other methods in the literature, may be applied
in a straightforward way to 
dynamical systems
with more than two degrees of freedom and is equally efficient for single
trajectories as well as large sets of them.
The method consists in analyzing a time series
constructed by the values of the geodesic deviation of nearby
trajectories recorded at a properly selected frequency. It should be
noted that a method based on a similar technique has been proposed by 
Lohinger and Froeschl\'{e} (1993).

The paper is organized as follows. Section 2 describes the basic
features of the method, which, in Section 3, is tested upon three
dynamical systems of different types. A comparison of the results of
our method to those derived by various other methods is made in Section
4. Section 5 treats the application of the proposed method to one of
the most important problems of solar system dynamics, the motion of
asteroids. Finally in Section 6 we present our conclusions.

\section{The method}

In order to decide on the nature of a trajectory (chaotic or not), we
work as follows. We integrate numerically the ``main" trajectory together with
a nearby one, which at a time $t_0=0$ starts at an infinitesimal
distance in phase space, $d_0$, from the main, and we calculate  their
distance, $d_1$, at a time $t = t_0 + \Delta t$. Let us denote by $q$ the
logarithm of the ratio of the two distances, $q = \ln (d_1/d_0)$. 
We then renormalize the
nearby trajectory, so as to start from a new position in phase space, which
is at distance $d_0$ from the main trajectory in the direction of
$d_1$. The trajectories are 
followed once again for a time interval $\Delta t$ and a new $q$ is
calculated. After following the trajectories for a time interval $T =
N~\Delta t$, we have constructed a time series consisting of the
consecutive $q$'s 
\begin{equation}
q(t) = \ln \left[{d_t \over d_0}\right]_t~~~ {\rm or}~~~
 q_k = \ln \left[{d_k \over d_0}\right]_{t_k}
\label{eq_qk}
\end{equation}
 
An ordered trajectory of a N-D conservative dynamical system will lie on an
invariant torus, i.e. a N-D 
manifold of the 2N-D phase space. Any such trajectory is, in general,
quasi-periodic and covers densely the 
invariant set. If a nearby trajectory is started at an infinitesimal
distance from  the previous one, this too would, in general, lie on an
invariant torus, so that the $q(t)$ time series should behave in a 
quasi-periodic manner. On the other hand, a chaotic trajectory visits
different regions of phase space in a stochastic manner and $q(t)$ should also 
be ``random''. 

Following the above considerations we calculate the {\it power
spectrum} $P(f)$ of the $q(t)$ time series. We first calculate
the discrete Fourier transform of the $q_k$ multiplied by a {\it
window function} $w_k$
\begin{equation}
Q_j=\sum_{k=0}^{N-1}q_k w_k e^{2\pi i j k /N} ~~~~~j=0..N-1
\end{equation}
Then the power spectrum, $P(f)$, is defined in $M=N/2+1$ frequencies as 
\begin{eqnarray}
P(f_0)&=&  \frac{1}{W} ~ \vert Q_0 \vert^2 \nonumber \\ 
P(f_j)&=&  \frac{1}{W} ~ \left( \vert Q_j \vert^2 + \vert Q_{N-j} \vert^2 
\right )~~~~~ j=1..(\frac{N}{2}-1) \\
P(f_c) &=&  \frac{1}{W}~ \vert Q_{N/2} \vert^2 \nonumber  
\end{eqnarray}
where
\begin{equation}
W  = N \sum_{k=0}^{N-1}w_k^2 
\end{equation}
and $f_c = f_{N/2}$ is the Nyquist frequency defined as
\begin{equation}
f_c=\frac{1}{2\Delta t}.
\end{equation}
The frequencies covered by the power spectrum are
\begin{equation}
f_j=f_c~\frac{j}{M}~~~~~~~j=0..M
\label{eq_freqs}
\end{equation}
In the present work we used the so called ``Hanning'' window. More details
on the calculation of the
power spectrum can be found in the book by Press et. al. (1992).

\begin{figure}
\resizebox{\hsize}{!}{\rotatebox{270}{\includegraphics*{9135f01a.ps}}
\hspace{1cm}          \rotatebox{270}{\includegraphics*{9135f01b.ps}}}
\vskip 0.1cm
\resizebox{\hsize}{!}{\rotatebox{270}{\includegraphics*{9135f01c.ps}}
\hspace{1cm}          \rotatebox{270}{\includegraphics*{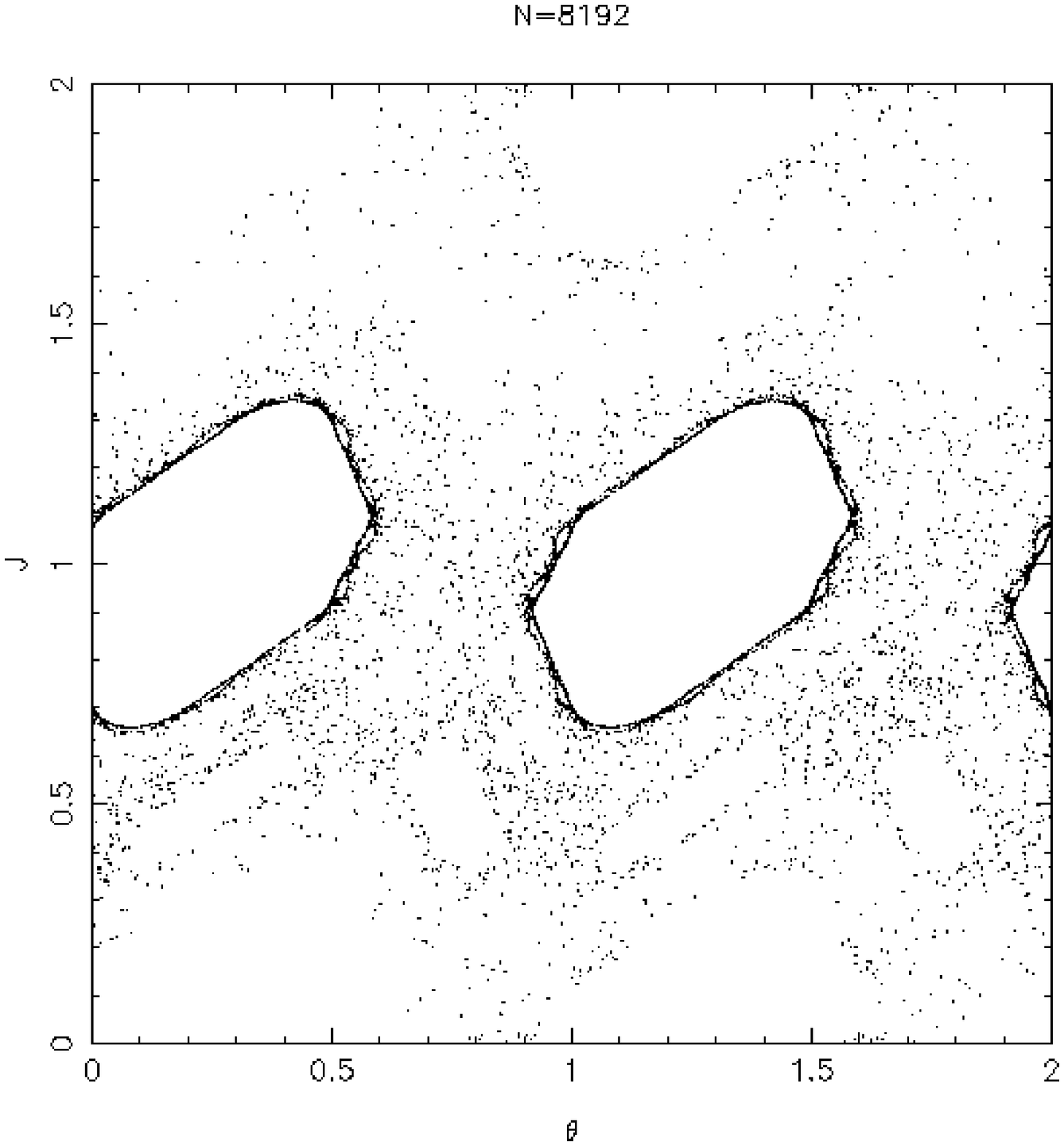}}}
\caption{The successive points of the ``sticky'' trajectory (starting at 
$J_0 = \pi $, $\theta_0 = 1.538~\pi$). Upper left, the first
2048 points, upper right the first 4096, down left, the first 6144
and down right, the first 8192. The units in all axes are given as
multiples of $\pi$ }
\label{map_invar}
\end{figure}

As a rule ``mixed" methods are expected to perform better in distinguishing 
between ordered and chaotic trajectories. The reason for that 
is that time series that are constructed from the geodesic divergence of
nearby orbits contain all the various characteristic frequencies that 
locally affect the motion in the ``proper" ratio, i.e. the frequencies 
corresponding to the different directions (degrees of freedom) are 
properly weighted.In particular now, as far as our method is concerned,
the Power Spectrum Of Divergences (PSOD) of an ordered
trajectory is expected to posses only a few ``spikes" at specific
frequencies. The number of harmonics however depends on the system under
consideration as well as on the values of its ``controll'' parameters 
(see below). 
In contrast, the PSOD of a chaotic trajectory should appear
continuous, due to the  random nature of the  $q_k$ time series.
However, the above considerations lie behind all methods based on time
series analysis. What is really important, for the assessment of the
new method with respect to the 
other ones appearing in the literature, is to evaluate (a) its
independence from the number 
of degrees of freedom of the dynamical system, (b) its effectiveness
with respect to the object of 
test (single trajectories or distributions of initial conditions
covering wide phase space regions) (c) its
sensitivity, i.e. the minimum length of the time series, necessary to 
distinguish a sticky chaotic trajectory from an ordered one and (d) its
ability to produce a well-defined measure of chaos (as is the LCN). 
 
\begin{figure}
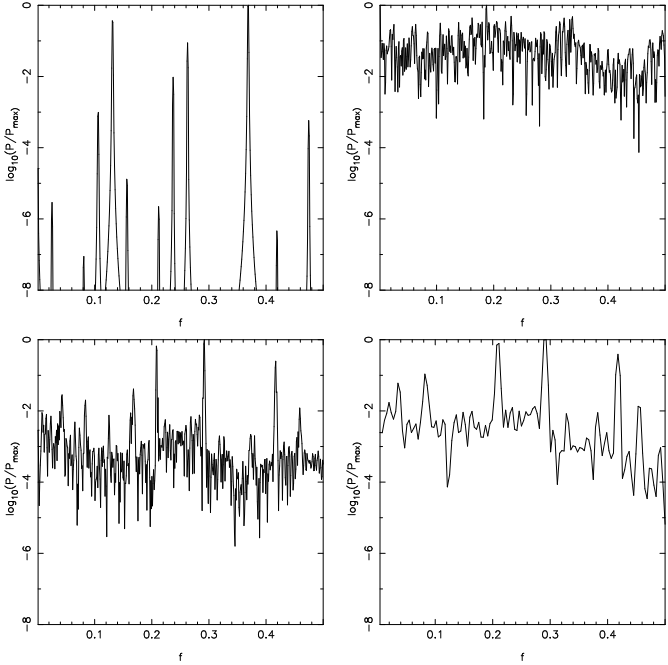

\resizebox{\hsize}{!}{\rotatebox{270}{\includegraphics*{9135f02a.ps}}
\hspace{1cm}          \rotatebox{270}{\includegraphics*{9135f02b.ps}}}
\vskip 0.1cm
\resizebox{\hsize}{!}{\rotatebox{270}{\includegraphics*{9135f02c.ps}}
\hspace{1cm}          \rotatebox{270}{\includegraphics*{9135f02d.ps}}}
\caption{The PSOD for the three trajectories of the 2-D mapping and 
for N=1024 iterations.
 Upper left : {\it map1} trajectory,
upper right : {\it map2} trajectory, down left : {\it map3} trajectory and
down right : {\it map3} trajectory but now with  N=256 }
\label{map_PSOD}
\end{figure}

\section{Evaluation of the method}

We proceed in the assessment of the method using three different
dynamical systems, namely a 
2-D mapping as well as a 2-D and a 3-D Hamiltonian system. In each one
we evaluated the nature of a considerable number of trajectories. In the
following subsections   
we present only three or four trajectories per system, which we think that
are typical examples of  
the three different classes of trajectories, i.e. ordered, clearly
chaotic and sticky. The amplitudes of the 
PSOD, in all figures, are normalized so that the highest has the value
one, while the frequency is given in cycles per time unit. 

\subsection{2-D mapping}

\begin{figure}
\resizebox{\hsize}{!}{\rotatebox{270}{\includegraphics*{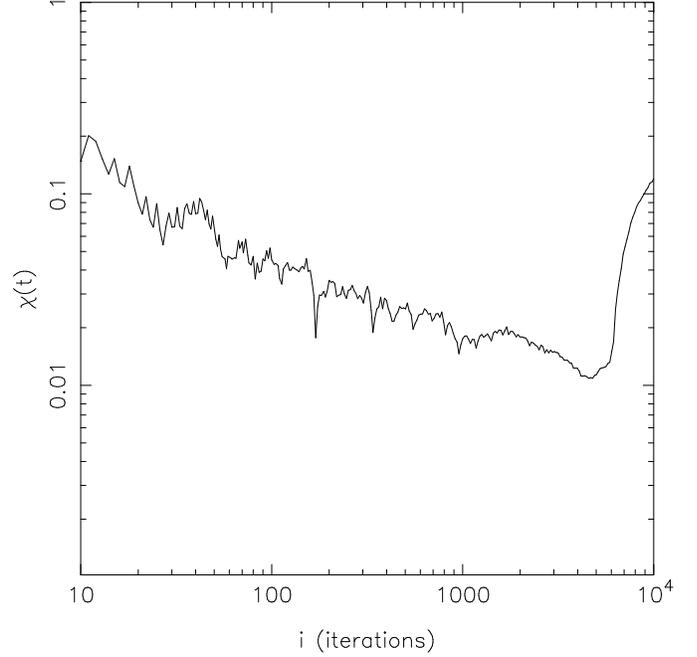}}}
\caption{The evolution of $\chi(t)$ for the sticky trajectory in the 2-D mapping }
\label{map_LCN}
\end{figure}

We first test the method in the simple 2-D mapping
\begin{eqnarray}
 J_{i+1} & = & J_i + k \cos(2 \theta_i)~~~\mathrm{mod}(2\pi) \nonumber\\ 
\theta_{i+1} & = & \theta_i + J_{i+1} ~~~~~~~~~~~~\mathrm{mod}(2\pi)
\end{eqnarray}
where the stochasticity parameter $k$ is taken equal to 0.7.

We present here the 
results of three trajectories, an ordered one ({\it map1}) starting at 
$J_0 = \pi $, $\theta_0 = 1.4~\pi$, 
a  stochastic one ({\it map2}) starting at 
$J_0 = \pi $, $\theta_0 = 1.5~\pi$ and a sticky one ({\it map3})
starting at 
$J_0 = \pi $, $\theta_0 = 1.538~\pi$. The initial conditions of the
third one place it very close to the boundary
between the ordered region, surrounding the stable point 
$J_0 = \pi $, $\theta_0 = 5\pi/4$, and the chaotic sea. Figure 
\ref{map_invar}
shows the consequents  ($\theta$, $J$) of the ``sticky'' trajectory at
various times (iterations). 
As we can see, for up to 2\,048 iterations the trajectory
behaves like an ordered one. Around $i=4\,000$ it starts to
present  some signs of irregularity and finally,
after $i=6\,000$,  the chaotic nature of the trajectory becomes evident.
Using our method on these three trajectories, with the length of the
$q_k$ time  
series being N=1024 iterations and defining $\Delta t = 1$, we obtain the 
spectra shown in Fig.~\ref{map_PSOD}. 

The upper left frame of Fig.~\ref{map_PSOD} is the spectrum of the ordered
trajectory. 
The spectrum consists of some basic frequencies while the
``noise'' is at a very low level. On the contrary, the spectrum
of the chaotic trajectory, in the upper right frame, covers the whole
frequency range with comparable amplitudes,
i.e. a clearly continuous spectrum. Looking at the spectrum of 
the ``sticky'' trajectory (lower left frame), we see a pattern almost
the same as that of the chaotic one. Some
high-amplitude spikes are evident  
but the continuum noise level is again very high for the whole frequency 
range, denoting the stochastic nature of the trajectory. Even with much less
iterations (N=256 - lower right frame of Fig.~\ref{map_PSOD}) we get
the same result. The spectrum is less dense, since it has a smaller 
number of frequencies than before (see eq.(\ref{eq_freqs})), but the 
basic features are the same.  Note that for the LCN, which is  the 
limit of the function 
\begin{equation}
\chi(t)= \frac{1}{N~\Delta t}\sum_{i=1}^{N} q_i
\end{equation}
as $N \rightarrow \infty$ (see Fig.~\ref{map_LCN}), or the plot 
of the consequents of the mapping (Fig.~\ref{map_invar}), we need much
more than 1024 iterations in order to decide whether the trajectory 
is chaotic or not. 

\subsection{An improved criterion}

\begin{figure}
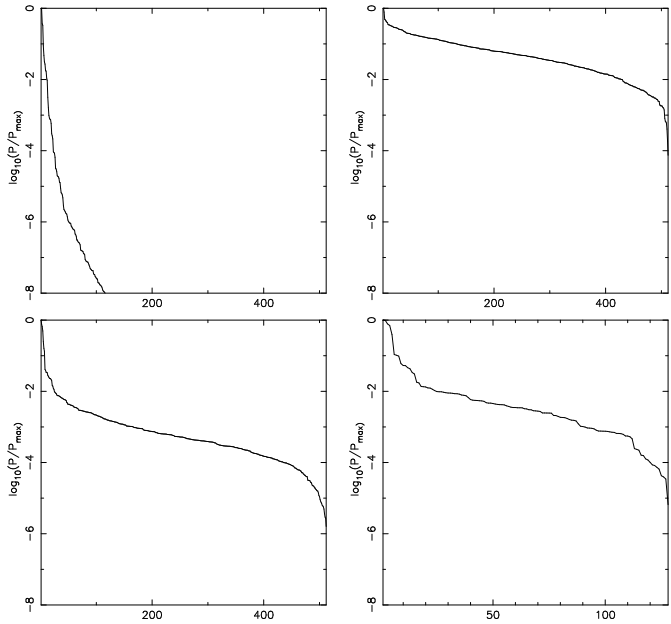

\resizebox{\hsize}{!}{\rotatebox{270}{\includegraphics*{9135f04a.ps}}
\hspace{1cm}          \rotatebox{270}{\includegraphics*{9135f04b.ps}}}
\vskip 0.1cm
\resizebox{\hsize}{!}{\rotatebox{270}{\includegraphics*{9135f04c.ps}}
\hspace{1cm}          \rotatebox{270}{\includegraphics*{9135f04d.ps}}}
\caption{The PSOD for the three 
trajectories in the 2-D mapping shown in Fig.~\ref{map_PSOD}, the peaks 
being sorted in descending order of amplitude.
The first three frames are for N=1024 iterations, while the fouth (
down right ) is {\it map3} trajectory with  N=256 iterations.}
\label{map_PSOD_sort}
\end{figure}

The above presented results show that it is, indeed, worth to consider
the new method as a useful tool in the assessment of the nature,
ordered or chaotic, of a trajectory. However the method, as it is,
does not entail a {\it clear} and {\it easy to apply} criterion for the
classification of a trajectory as ordered or chaotic. Here we try to
improve somehow the presentation of the results of our method, in order 
to propose such a criterion. Note that this new criterion is 
similar, graphically, to the criterion of the
FLI proposed by Froeschl\'e et al (1997).

If the peaks appearing in the PSOD are plotted in descending order of 
amplitude, we have a graphical representation of how many strong 
frequencies the spectrum possesses. Figure~\ref{map_PSOD_sort} shows this 
representation of the PSOD for the three
trajectories studied in the previous sub-section. Ordered trajectories
have only a few high amplitude frequencies
and the background is formed by peaks whose amplitudes are more than four 
orders of magnitude smaller than that of the basic frequency.
On the other hand the stochastic trajectories 
possess only a few high-amplitude frequencies and the largest part of the
spectrum consists of a ``continuum'' of frequencies with also considerable
amplitude. In this respect one may chose to represent the results by a signle 
number (e.g. the number of peaks up to a certain amplitude), provided that
a certain ``threshold'' for the noise level is chosen (10$^{-8}$ in Fig. 
4). This value will, however, depend on the system at study.

\subsection{Noise level}
\label{section_noise}

\begin{figure}
\resizebox{\hsize}{!}{\rotatebox{270}{\includegraphics*{9135f05a.ps}}
\hspace{1cm}          \rotatebox{270}{\includegraphics*{9135f05b.ps}}}
\vskip 0.1cm
\resizebox{\hsize}{!}{\rotatebox{270}{\includegraphics*{9135f05c.ps}}
\hspace{1cm}          \rotatebox{270}{\includegraphics*{9135f05d.ps}}}
\vskip 0.1cm
\resizebox{\hsize}{!}{\rotatebox{270}{\includegraphics*{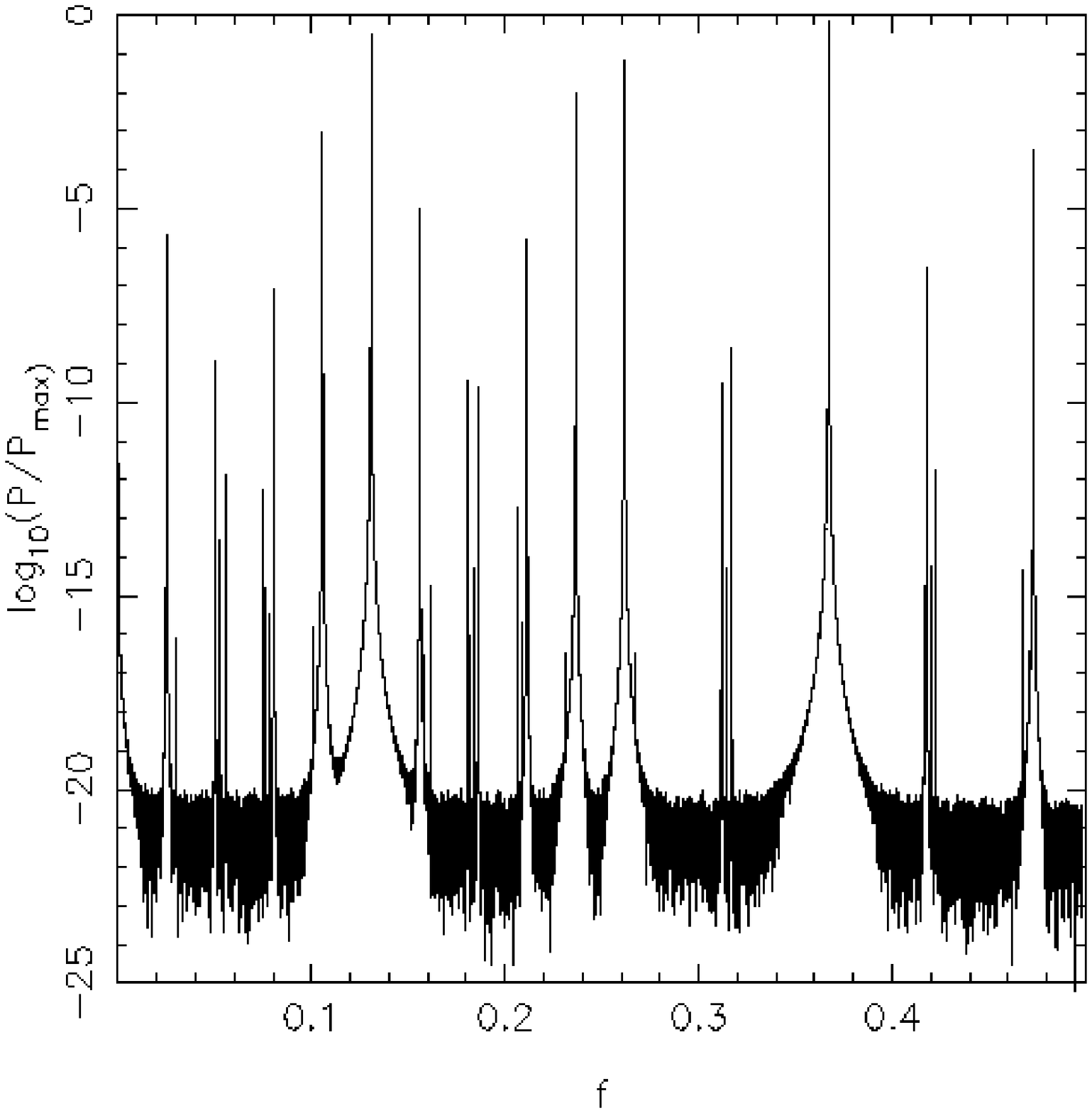}}
\hspace{1cm}          \rotatebox{270}{\includegraphics*{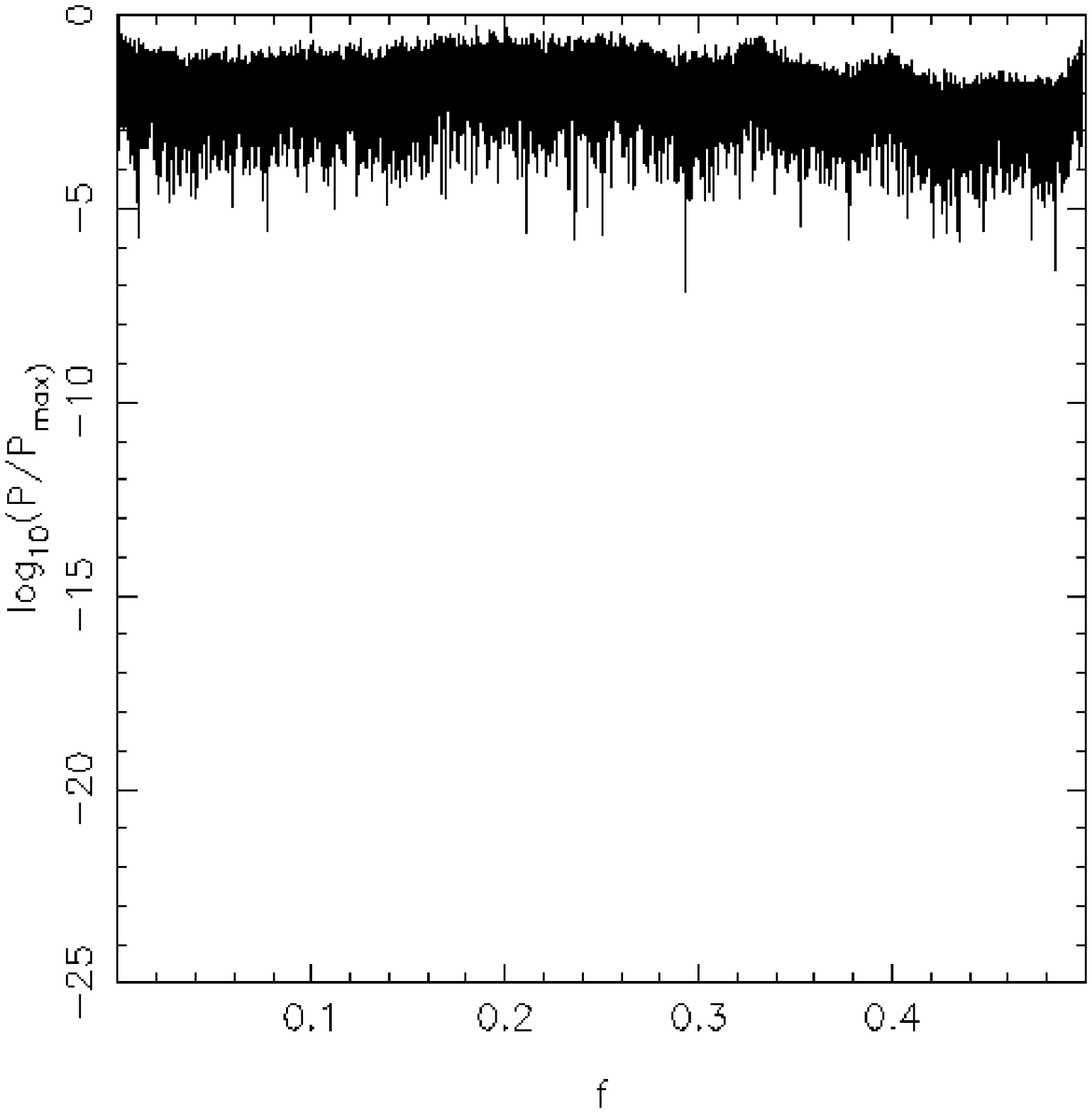}}}
\caption{The PSOD for the regular (left) and the chaotic (right)  
trajectories in the 2-D mapping with N=256 (top), N=4096 (middle) and
N=65\,536 (bottom).}
\label{noise1}
\end{figure}

\begin{figure}
\resizebox{\hsize}{!}{\rotatebox{270}{\includegraphics*{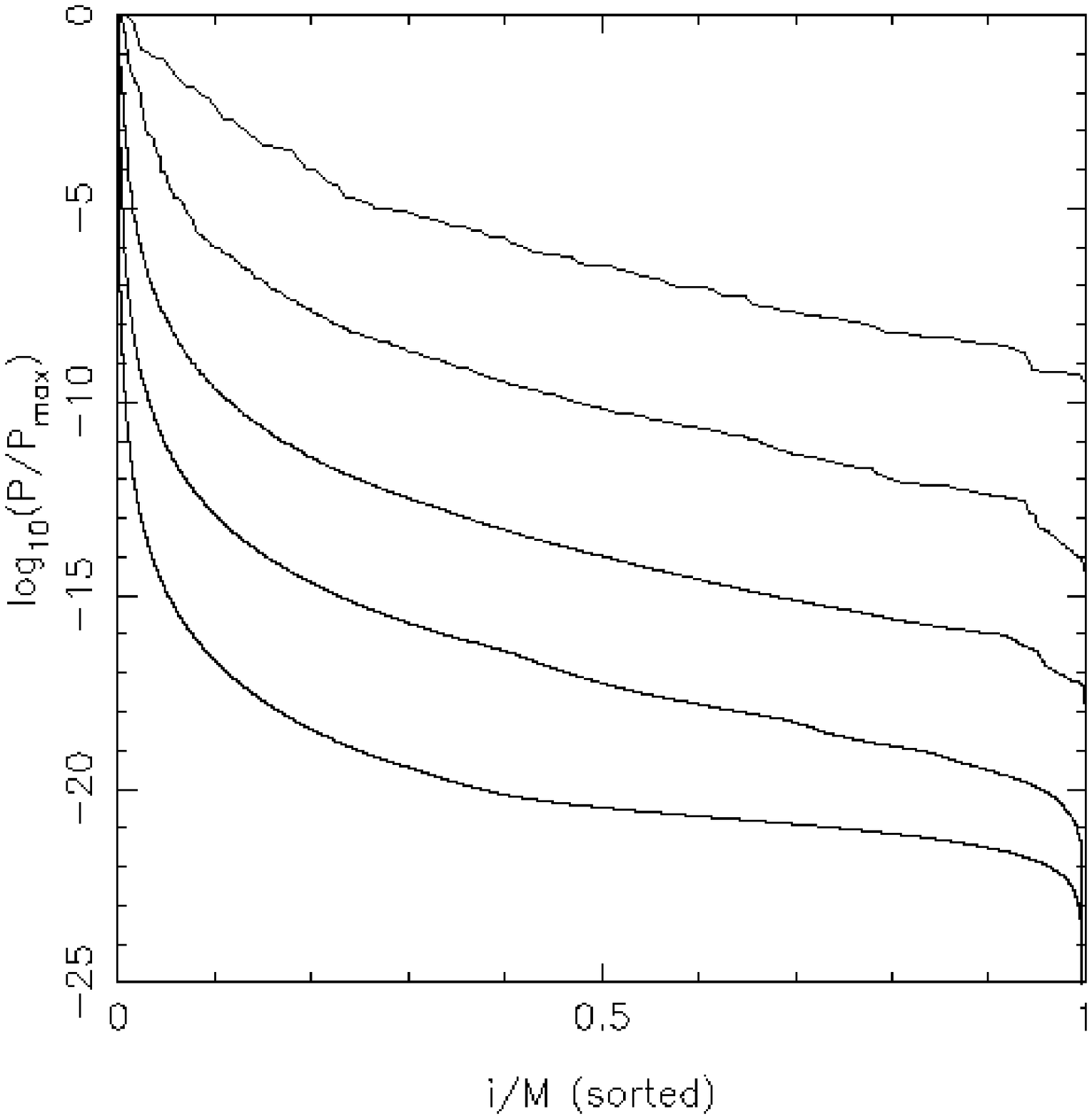}}
\hspace{1cm}          \rotatebox{270}{\includegraphics*{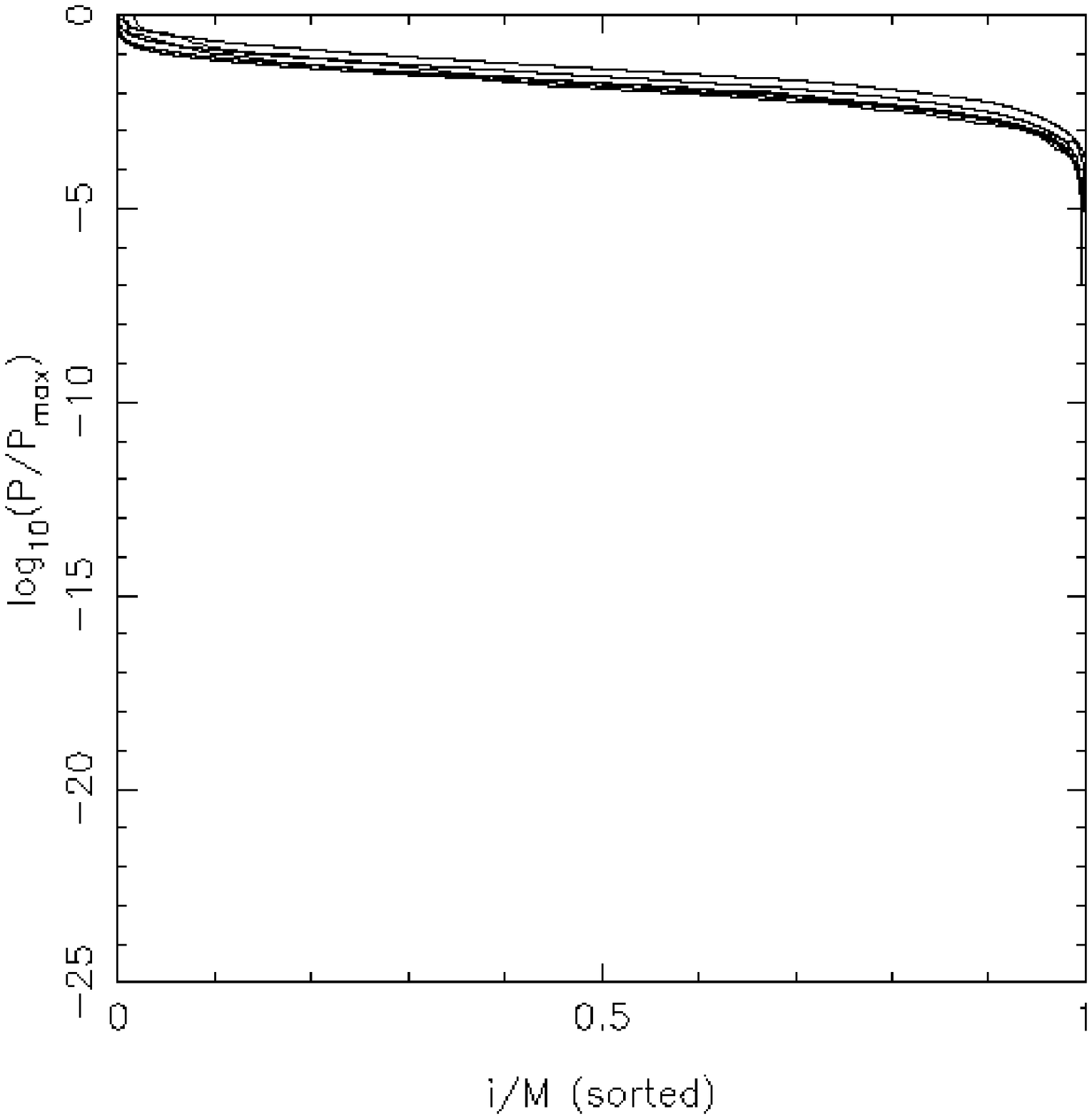}}}
\caption{The sorted PSOD for the regular (left) and the chaotic (right)  
trajectories in the 2-D mapping for different iteration number, N. Lines 
from top to bottom correspond to N=256, 1024, 4096, 16\,384 and  65\,536. }
\label{noise2}
\end{figure}

Any method of analysis of finite-sample time series is bound to suffer 
from noise. Thus, even in the PSOD of a regular orbit a certain noise 
level is expected. This is mainly due to ``leakage'' of power from 
the frequency lobes, a side effect of the calculation of the power 
spectrum using Discrete Fast Fourier Transform methods. When calculating 
the power spectrum of a ``monochromatic'' signal, the power contained in 
its basic frequency ``leaks'' into neighboring frequencies. The leakage 
depends on the windowing function used. When a signal possesses two 
closely spaced frequencies, all frequencies in between these two will 
also gain considerable amplitudes, due to this phenomenon.
If the PSOD of a regular orbit has a large number of basic (strong) 
frequencies, then this effect can lead us to falsely identify it as chaotic. 
The problem can be tackled at the cost of taking more points in the sample. 
In this way, the number of frequencies appearing in the spectrum is 
increased but the frequency lobes become thinner. Thus, lobe overlapping is 
reduced and the noise level drops.
For chaotic orbits on the other hand, the observed noisy pattern
is an inherent property of the spectrum and by increasing the number of 
points one cannot alter the picture.

The above can be seen in Fig.~\ref{noise1}, which shows the PSOD of a 
regular (left column) and a
chaotic trajectory (right column) for three different values of N, i.e. 
N=256 (top), N=4096 (middle) and N=65\,536 (bottom). As we see, the noise
level in the case of the regular orbit drops significantly when N
is increased, while in the case of the chaotic orbit it remains more or
less unchanged. In the N=65\,536 case the noise level for the ordered 
trajectory drops below $10^{-20}$, becoming comparable to the 
accuracy of the FFT calculation (double precision).

This phenomenon is better seen if we use the amplitude-sorted
PSOD. Fig.~\ref{noise2} shows the amplitude-sorted PSOD of the regular 
(left) and chaotic (right) trajectory with N=256, 1024, 4096, 16\,384 and
65\,536. While the noise level in the PSOD of the ordered trajectory is 
reduced when N is increased, it remains the same for the chaotic one.

\subsection{2-D Hamiltonian system}

\begin{figure}
\resizebox{\hsize}{!}{\rotatebox{270}{\includegraphics*{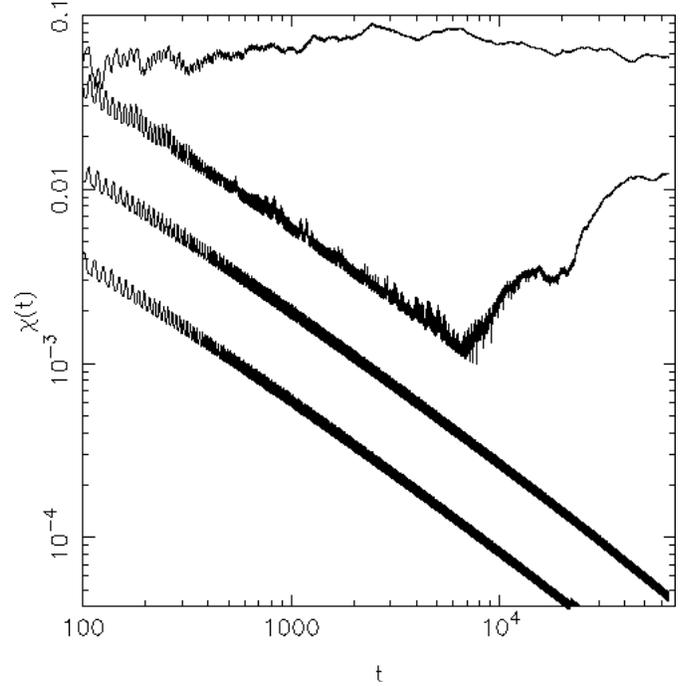}}}
\caption{The evolution of $\chi(t)$ of the four test trajectories in
the 2-D Hamiltonian model in the case $a=2$. The initial positions of
the trajectories are given in the text. Note that the $\chi(t)$ curves
of the two ordered trajectories have been lowered by one ({\it ord-2})
and half ({\it ord-2a)} orders of
magnitude in order to be clearly distinguished from the sticky trajectory}
\label{2d_a2_LCN}
\end{figure}

\begin{figure}
\resizebox{\hsize}{!}{\rotatebox{270}{\includegraphics*{9135f08a.ps}}
\hspace{1cm}          \rotatebox{270}{\includegraphics*{9135f08b.ps}}}
\vskip 0.1cm
\resizebox{\hsize}{!}{\rotatebox{270}{\includegraphics*{9135f08c.ps}}
\hspace{1cm}          \rotatebox{270}{\includegraphics*{9135f08d.ps}}}
\caption{The PSOD, with N=2048, of the four test trajectories in the
2-D Hamiltonian model for $a$ = 2. Upper left: the ordered {\it
ord-2}, upper right:  the chaotic {\it ch-2}, lower left: the
sticky {\it st-2} and lower right: the ordered {\it ord-2a}. }
\label{2d_a2_PSOD}
\end{figure}

\begin{figure}
\resizebox{\hsize}{!}{\rotatebox{270}{\includegraphics*{9135f09a.ps}}
\hspace{1cm}          \rotatebox{270}{\includegraphics*{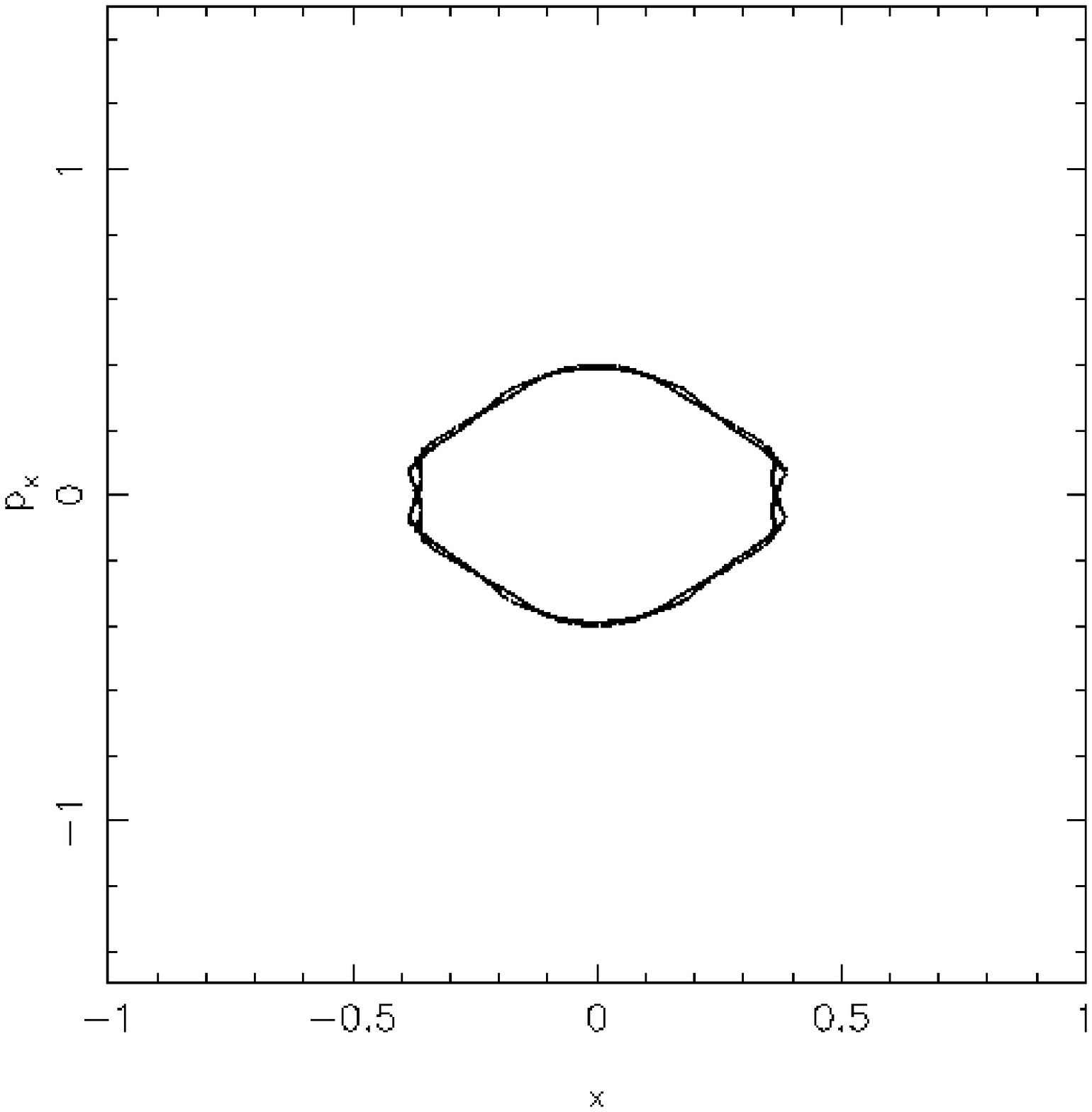}}}
\vskip 0.1cm
\resizebox{\hsize}{!}{\rotatebox{270}{\includegraphics*{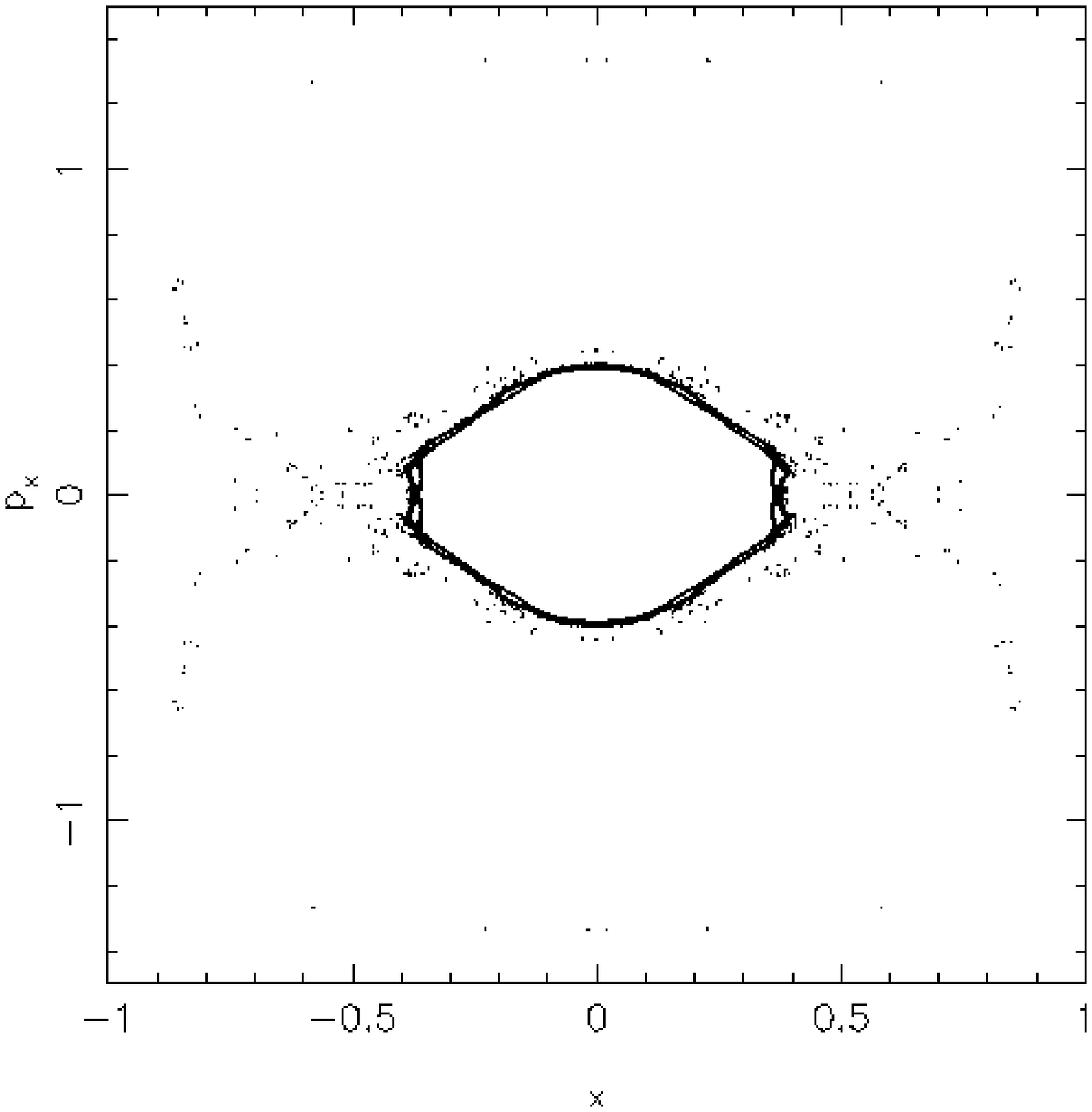}}
\hspace{1cm}          \rotatebox{270}{\includegraphics*{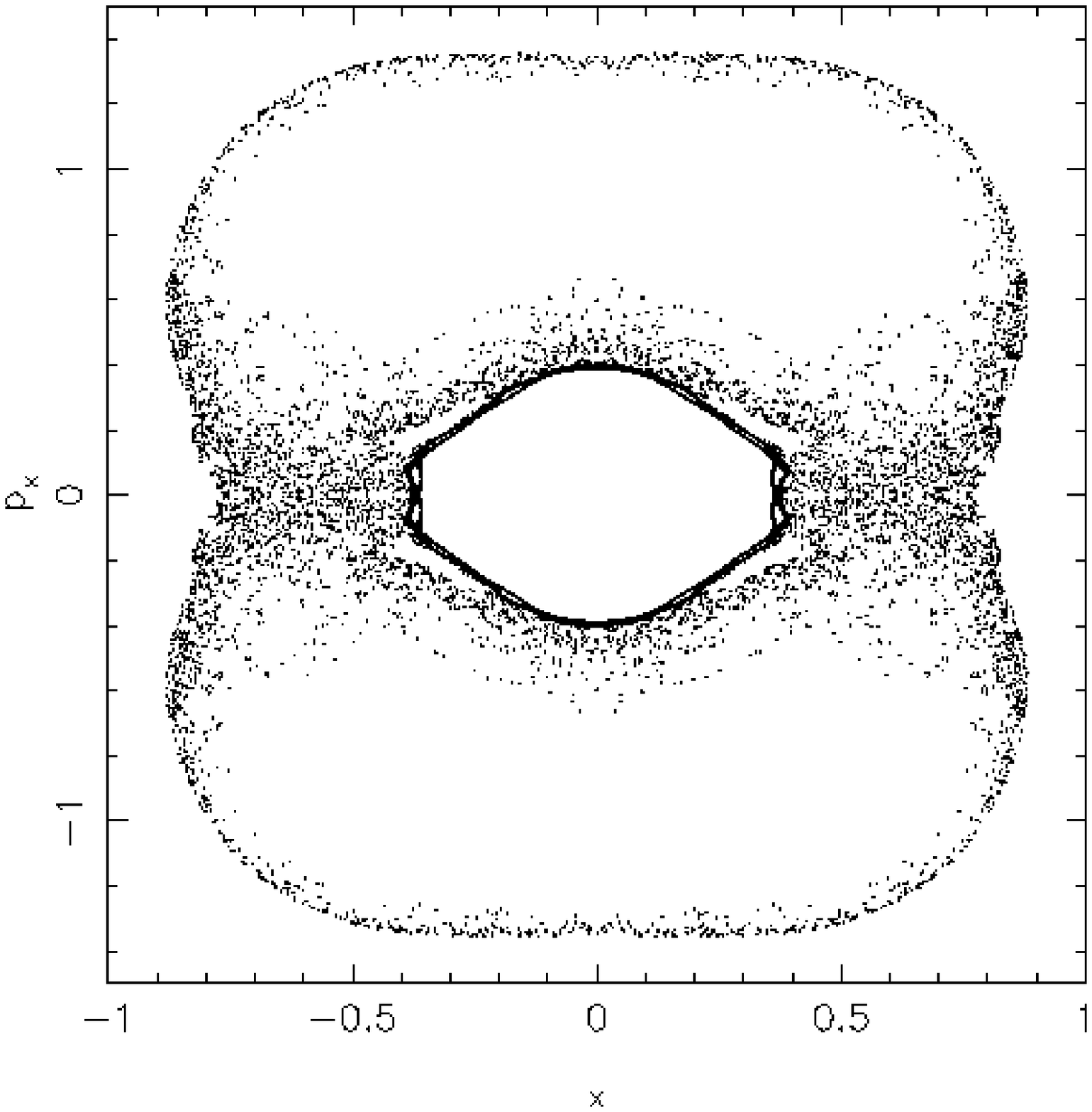}}}
\caption{The surface of section $x$, $\dot x$ of {\it st-4} in 
the 2-D Hamiltonian model. Upper left $t=3\,000$, upper right $t=8\,000$,
down left $t=15\,000$, down right $t=20\,000$}
\label{2d_inv}
\end{figure}

\begin{figure}
\resizebox{\hsize}{!}{\rotatebox{270}{\includegraphics*{9135f10a.ps}}
\hspace{1cm}          \rotatebox{270}{\includegraphics*{9135f10b.ps}}}
\vskip 0.1cm
\resizebox{\hsize}{!}{\rotatebox{270}{\includegraphics*{9135f10c.ps}}
\hspace{1cm}          \rotatebox{270}{\includegraphics*{9135f10d.ps}}}
\caption{The PSOD, with N=4096, of the four test trajectories in the
2-D Hamiltonian model for $a$ = 4. Upper left: the ordered {\it
ord-4}, upper right:  the chaotic {\it ch-4} and lower left: the
sticky {\it st-4} and lower right: the ordered {\it
ord-4a}. }
\label{2d_PSOD}
\end{figure}

\begin{figure}
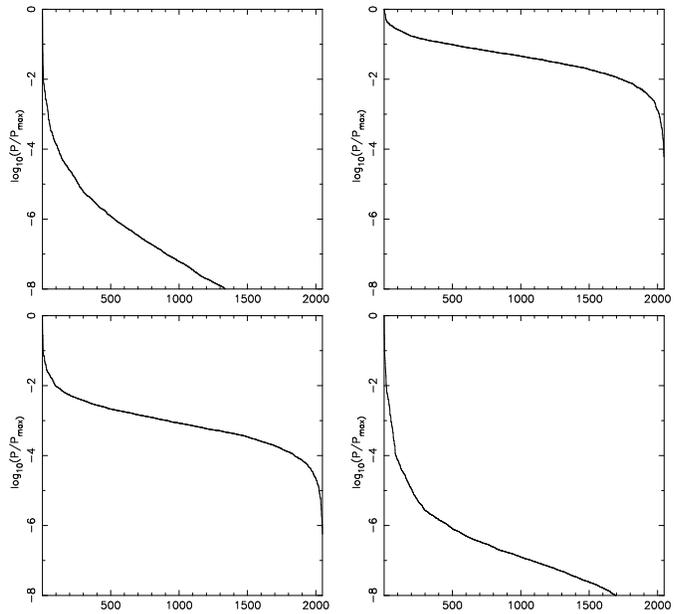

\resizebox{\hsize}{!}{\rotatebox{270}{\includegraphics*{9135f11a.ps}} 
\hspace{1cm}          \rotatebox{270}{\includegraphics*{9135f11b.ps}}}
\vskip 0.1cm
\resizebox{\hsize}{!}{\rotatebox{270}{\includegraphics*{9135f11c.ps}}
\hspace{1cm}          \rotatebox{270}{\includegraphics*{9135f11d.ps}}} 
\caption{The sorted PSOD of Fig.~\ref{2d_PSOD} }
\label{2d_PSOD_sort}
\end{figure}

We now proceed to test the method in a 2-D Hamiltonian system.
We selected the Hamiltonian used by Caranicolas \& Vozikis (1987),
\begin{equation}
 H = {1 \over 2} \left( p_x^2 + p_y^2 \right) + x^4 + y^4 + 2
a x^2y^2 = h
\end{equation}
where the parameter $a$ is taken equal to 2.0 and the energy constant
$h=1.0$.

Again we  study three trajectories, one ordered ({\it ord-2}) starting at
$x(0) = 0.5$, one clearly stochastic ({\it ch-2}) starting at $x(0) = 0.001$ 
and one ``sticky'' ({\it st-2}) which starts at  $x(0) = 0.1326$.
All three trajectories have also  $y(0) = 0$ and $p_x(0) = 0$, while their
$p_y(0)$ is given by the energy integral.

Using the PSOD with a renormalization time-step equal to $\Delta t = 1.5$,
we find that the spectra of the three test trajectories 
present the same properties as those of the corresponding cases of the 
mapping. In Fig.~\ref{2d_a2_PSOD} we see the PSOD of the three test 
trajectories. The difference between the spectrum of the
chaotic trajectory {\it ch-2} and that of the ordered trajectory {\it ord-2}
is again obvious. Moreover, the ``sticky'' trajectory {\it st-2}
has a spectrum similar to that of {\it ch-2}. Note that we have used only
2048 points (corresponding to $t= 3072$) while, if we look at the
evolution of the $\chi(t)$ (Fig.~\ref{2d_a2_LCN}), the trajectory 
looks ordered for a time up to $t\approx 7\,000$. The fourth frame 
(lower right) in Fig.~\ref{2d_a2_PSOD} corresponds to another regular orbit 
({\it ord-2a}) that starts at $x= 0.133$, i.e. very close to the sticky 
trajectory. We see that, although the two trajectories start very
close to each other and, at least for the first 3\,000 times steps,
span approximately the same phase-space region, their PSODs are completely
different, clearly revealing the nature of each case. 

We decided to test also the case where the parameter $a$ is taken
equal to $4.0$. As Caranicolas \& Vozikis (1987) have shown, the 
surface of section of this case has a completely different topology 
from that of the $a=2$ case. The equipotential curves have negative
curvature along the $y=\pm x$ lines. This affects mainly the loop
orbits which appear ``squared''.
 
Again we  study four trajectories, one ordered ({\it ord-4}) starting at
$x(0) = 0.3$, one clearly chaotic ({\it ch-4}) starting at $x(0) = 0.7$, 
one ``sticky'' ({\it st-4}) which starts at  $x(0) = 0.36282$ and one
ordered ({\it ord-4a}) starting  at $x(0)= 0.362$ very near to the sticky 
one. The surface of section plot for the ``sticky'' trajectory is shown in 
Fig.~\ref{2d_inv} at various times. Figures~\ref{2d_PSOD} and 
\ref{2d_PSOD_sort} present the PSOD and the amplitude-sorted PSOD for 
these four orbits (N=4096). An important characteristic seen in these two 
figures is the presence of a high number of medium-amplitude frequencies 
in the spectra of the regular orbits. However, a distinction between regular
and chaotic orbits can still be made. Due to frequency overlapping, 
discussed in section \ref{section_noise}, individual frequencies cannot 
be distinguished at an amplitude level smaller than $10^{-6}$. This makes 
it very difficult to identify a sticky orbit with an amplitude level 
around $10^{-6}-10^{-7}$. Nevertheless the noise level of the regular 
orbits will be supressed if we take more points, while for a sticky 
chaotic orbit it will remain more or less the same.
 
\subsection{3-D Hamiltonian system}

We apply our method to the model Hamiltonian used by
Magnenat (1982), Contopoulos \& Barbanis (1989),
Barbanis and Contopoulos (1995), Barbanis (1996), Varvoglis et
al. (1997), Barbanis et. al. (1999) and Tsiganis et al. (2000a)
\begin{eqnarray}
H &=& {1 \over 2} \left( p_x^2 + p_y^2 + p_z^2 \right) \nonumber \\
&&+ {1 \over 2} \left( A x^2 + B y^2 + C z^2 \right) - \epsilon x z^2 -  
\eta y z^2  = h
\end{eqnarray}
where the parameters are taken as $A = 0.9$, $B = 0.4$, $C =
0.225$, $\epsilon = 0.560$ and $\eta = 0.20$ and the energy
level is $h = 0.00765$
               
\begin{figure}
\resizebox{\hsize}{!}{\rotatebox{270}{\includegraphics*{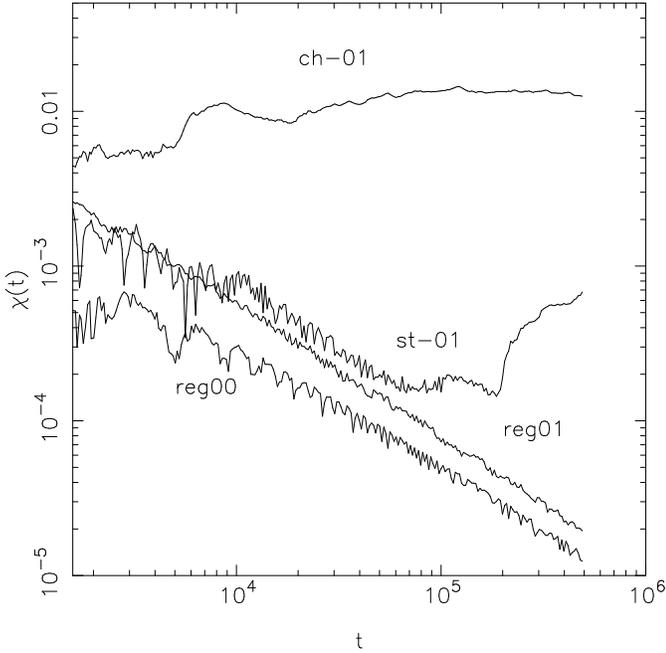}}}
\caption{The evolution of $\chi(t)$ of the four test trajectories in
the 3-D Hamiltonian 
model. The initial positions of the trajectories are given in the text.}
\label{3d1}
\end{figure}

We again test three trajectories, an ordered one ({\it reg01}) starting at
$\bar x =0.01$, $\bar y=0.032$, a chaotic one ({\it ch-01}) starting
at $\bar x=0$,  $\bar y=0$ and a sticky one
({\it st-01}) with initial conditions
$\bar x =0.01725$, $\bar y=0.032$, while $\bar z$, $p_x$, $p_y$
were taken equal to 0 and $p_z$ is calculated from the energy integral.
We use the variables $\bar x$, $\bar y$, $\bar z$ instead of $x$, $y$,
$z$ in order to be consistent with the previous publications.
The barred variables are defined as $\bar x= \sqrt{A}x$, $\bar y =
\sqrt{B} y$ and $\bar z = \sqrt{C} z$.

In 3-D one cannot visualize a surface of section plot, in order to check 
whether a
particular trajectory is ordered or chaotic. Therefore, if one is using
the traditional tools, he has to rely on the calculation of LCNs. It 
should be pointed out that a positive LCN is a proof that the trajectory
under study is chaotic, while a monotonically decreasing value of $\chi
(t)$ is not a proof of order, since this behavior could very well 
originate from ``stickiness". That is why we decided to test one 
more trajectory ({\it reg00}) for which we can be almost certain that it is 
ordered, as it has the same initial position as {\it reg01} but it belongs 
to an almost integrable case of the model Hamiltonian, $\epsilon=0.01$ and 
$\eta=0.01$.

Figure~\ref{3d1} shows the calculation of the $\chi(t)$ for the four
trajectories. We can clearly see 
that $\chi (t)$ of {\it st-01} is decreasing up to $t=7\,10^4$ and then
begins to saturate to a non-zero LCN value. The
stochasticity is even more evident after $t=2.2\,10^5$,
where we have a  ``jump'' to a higher LCN value. 

\begin{figure}
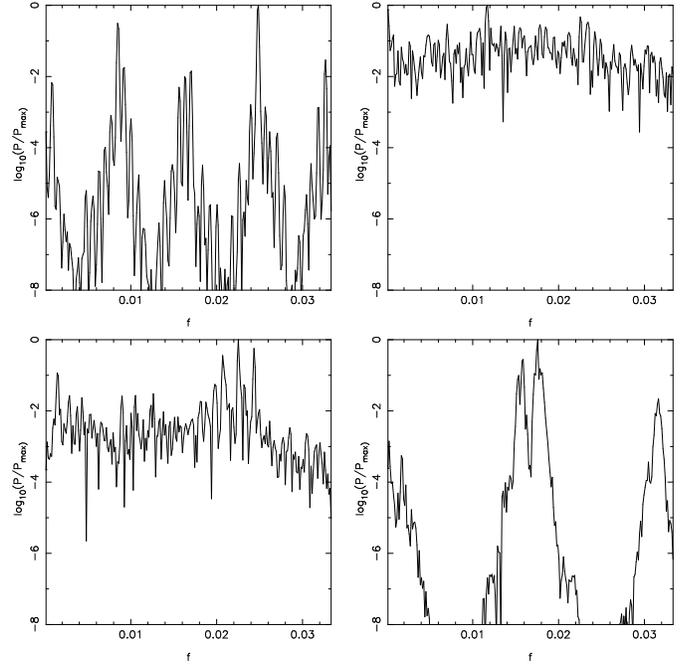

\resizebox{\hsize}{!}{\rotatebox{270}{\includegraphics*{9135f13a.ps}}
\hspace{1cm}          \rotatebox{270}{\includegraphics*{9135f13b.ps}}}
\vskip 0.1cm
\resizebox{\hsize}{!}{\rotatebox{270}{\includegraphics*{9135f13c.ps}}
\hspace{1cm}          \rotatebox{270}{\includegraphics*{9135f13d.ps}}}
\caption{The PSOD of the four test trajectories in the 3-D Hamiltonian
model using 512 points. i.e. 7680 time steps. Upper left the ordered
{\it reg01}, upper right the chaotic {\it ch-01}, lower left the
sticky {\it st-01} and lower right
the ordered {\it reg00}.}
\label{3d2}
\end{figure}

We calculate the PSOD using $\Delta t = 15$, a value approximately
equal to the time interval between two consecutive crossings of the 
$\bar x - \bar y$ plane by the trajectory.
Figure~\ref{3d2} shows the PSOD of the four trajectories using
512 points, i.e. for $t=7\,680$.
Note once again that we can decide that the ``sticky'' trajectory {\it st-01}
is actually stochastic well in advance of the LCN method.
The LCN shows the stochastic
behaviour of the trajectory only after $t=70\,000$, while with the PSOD we 
need a modest $t=7\,680$.

\begin{figure}
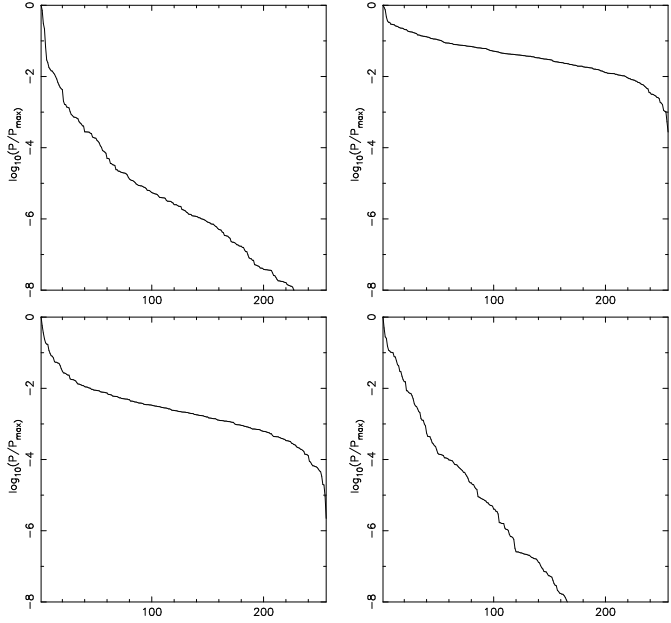

\resizebox{\hsize}{!}{\rotatebox{270}{\includegraphics*{9135f14a.ps}}
\hspace{1cm}          \rotatebox{270}{\includegraphics*{9135f14b.ps}}}
\vskip 0.1cm
\resizebox{\hsize}{!}{\rotatebox{270}{\includegraphics*{9135f14c.ps}}
\hspace{1cm}          \rotatebox{270}{\includegraphics*{9135f14d.ps}}}
\caption{The PSOD of the four test trajectories in the 3-D Hamiltonian
model shown in Fig.~\ref{3d2}, the peaks being sorted in descending order 
of amplitude.} 
\label{3d3}
\end{figure}

\section{Comparison with other methods}

As we already mentioned in the introduction, three of the most recent
methods for distinguishing ordered from chaotic trajectories are the Fast 
Lyapunov Indicators (FLI) (Froeschl\'e et al. 1997), the ``spectra'' of
stretching numbers and/or twist angles (Froeschl\'e et al. 1993, Voglis 
\& Contopoulos 1994, Contopoulos \& Voglis 1997) and the ``spectral 
distance'' (DSD) (Voglis et al. 1998, 1999). Out of these three methods 
the fastest ones are the FLI and the DSD methods. In this 
section we compare the PSOD with the FLI and the DSD methods by applying
it to the test models presented in the above mentioned papers. A 
discussion concerning the spectra of stretching numbers follows.

\begin{figure}
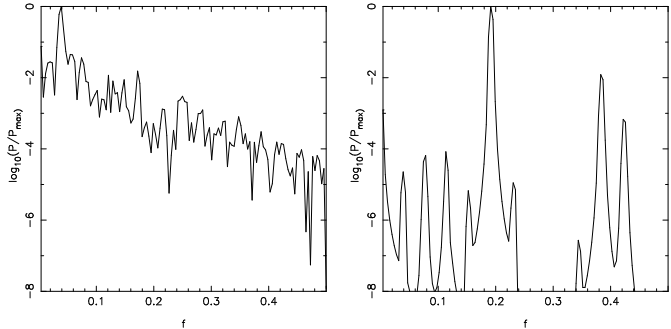

\resizebox{\hsize}{!}{\rotatebox{270}{\includegraphics*{9135f15a.ps}}
\hspace{1cm}          \rotatebox{270}{\includegraphics*{9135f15b.ps}}}
\caption{The PSOD of the two trajectories
used by Froeschl\'e et al. (1997) to test the FLI method. In the 
left frame is the PSOD of the chaotic trajectory ($x(0)=0.001$, $y(0)
=0.001$) and in the right frame is the PSOD of the ordered trajectory 
($x(0)=1$, $y(0)=0$).} 
\label{flg-psod}
\end{figure}

\subsection{Single trajectories}
\subsubsection{FLI}

In the paper by Froeschl\'e et al. (1997) the authors tested the FLI method
on two trajectories of the standard map
\begin{eqnarray}
x_{i+1}&=& x_i + k \sin(x_i + y_i) ~~ \mathrm{mod}(2\pi) \nonumber \\
y_{i+1}&=& x_i + y_i~~~~~~~~~~~~~~~~~~~~~~ \mathrm{mod}(2\pi) 
\end{eqnarray}
with $k=0.3$; one stochastic starting at
$x=0.001$, $y=0.001$ and one ordered starting at $x=1$, $y=0$. They
found that for the stochastic trajectory the FLI's drop very quickly down
to $10^{-20}$ (Fig.~2 in their paper - 200 iterations). On the contrary, 
the function $\chi (t)$ levels only after about 10\,000
iterations. For the ordered trajectory the FLI's are slowly decreasing,
following $\chi (t)$. In Fig.~\ref{flg-psod} we
present the PSOD's of the stochastic (left frame) and the ordered
(right frame) trajectory,
calculated using only 256 iterations.  It is obvious that the two
spectra clearly differentiate between the two types of trajectory.
  
\subsubsection{Spectral distance (the 4D map)}
\begin{figure}
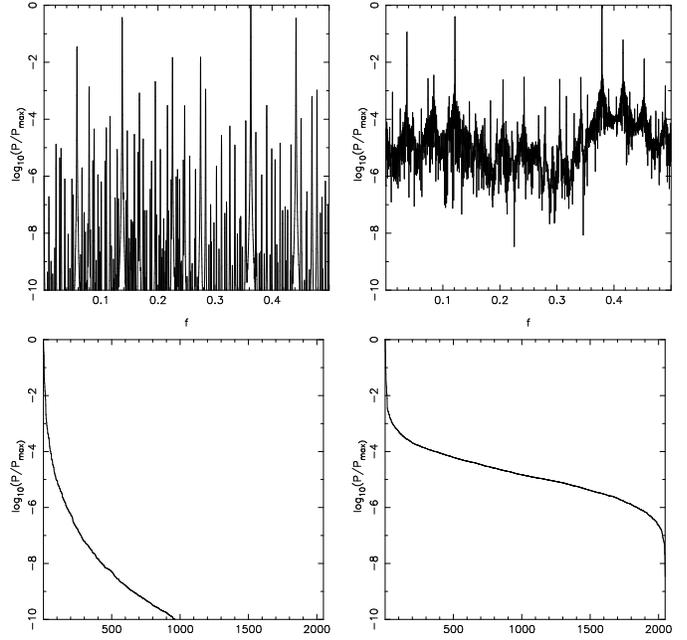

\resizebox{\hsize}{!}{\rotatebox{270}{\includegraphics*{9135f16a.ps}}
\hspace{1cm}          \rotatebox{270}{\includegraphics*{9135f16b.ps}}}
\vskip 0.1cm
\resizebox{\hsize}{!}{\rotatebox{270}{\includegraphics*{9135f16c.ps}}
\hspace{1cm}          \rotatebox{270}{\includegraphics*{9135f16d.ps}}}
\caption{The PSOD (upper frames) and the sorted PSOD (lower frames) 
of the two test trajectories (left frames : regular A2, right frames
: chaotic A3) in the 4-D mapping for N=4096.} 
\label{map4d}
\end{figure}

In a recent paper Voglis et al. (1999) proposed, as a tool for the
distinction between chaotic and regular orbits in 4D maps, the use of the 
``spectral distance'' $D^2$.
The method is based on the property that the ``spectrum'' of
{\it stretching numbers} (as well as that of the {\it helicity angles}) 
of a 
chaotic trajectory is independent of the initial orientation of the 
deviation vector, while the spectrum of a regular trajectory is not.

The ``spectral distance'' $D^2$ is a norm defined as
\begin{equation}
D^2 = \sum_q \left[ S_1(q) - S_2(q) \right]^2
\end{equation}
where the summation is for all $q$'s and $S_1$, $S_2$ are two spectra
of the same orbit but with two different initial deviation vectors.

Voglis et al. (1999) applied their method to a 4-D mapping
consisting of two coupled 2-D standard maps, i.e.
\begin{eqnarray}
x_1'&=&x_1+x_2' \nonumber\\
x_2'&=&x_2+\frac{k}{2\pi}\sin(2\pi
x_1)-\frac{\beta}{\pi}\sin(2\pi(x_1-x_3))\\ 
x_3'&=&x_3+x_4'\nonumber \\
x_4'&=&x_4+\frac{k}{2\pi}\sin(2\pi
x_3)-\frac{\beta}{\pi}\sin(2\pi(x_3-x_1))\nonumber 
\end{eqnarray}
where the $x_i$'s are defined in the interval [0,1) (i.e.$\mathrm{mod} 1$).

We chose to test our method upon the two most interesting cases shown in
Voglis et al. (1999), namely trajectories A2 and A3, in their notation. 
The A2 case has initial conditions ($x_1,x_2,
x_3,x_4$)=(0.55,0.1,0.62,0.2), $\beta=0.1$ and is a regular orbit, while 
the A3 case has the same initial $x_i$'s, $\beta=0.3051$ and is a chaotic 
orbit but with a very low value of LCN (around $4\,10^{-7}$).

Fig.\ref{map4d} shows the PSOD of the two test trajectories. The left
panel corresponds to the regular orbit (A2) and the right panel 
corresponds to the chaotic one (A3). The spectra were calculated using 4096 
iterations for each orbit. The distinction between the regular and the 
chaotic orbits is apparent. Note that our method gave the same result as the 
$D^2$ method with almost the same computational effort. Of course, both 
methods are much faster than the traditional LCN method.  

\subsubsection{``Spectrum'' of stretching numbers}
The ``spectrum'' of stretching numbers, $S(q)$, (Froeschl\'e et
al. 1993, Voglis \& Contopoulos 1994, Contopoulos \& Voglis 1997,
Dvorak et al. 1998) is a method also based  on the divergence of
nearby trajectories. 
It consists of the calculation of the probability density of 
$q_k$ (eq.\~(\ref{eq_qk})), i.e. 
\begin{equation}
S(q)=\frac{\Delta N(q)}{N dq}
\end{equation}
where $\Delta N(q)$ is the number of $q_k$ with values between $q$ and
$q+dq$. A quasi-periodic trajectory, which lies very close to a
periodic trajectory, has a ``U'' shaped distribution of stretching
numbers which is also symmetric around $q=0$ (Contopoulos et
al. 1997). As we move away from the periodic trajectory, this symmetry is
destroyed (Caranicolas \& Vozikis 1999) and the spectrum starts to
develop a greater number of maxima. On the 
other hand if the trajectory is chaotic, the spectra have different shapes
and are not symmetric at all.
It should be pointed out, however, that, in order to obtain a well defined spectrum, one
needs to account for a large number of iterations (typically $N=10^5$ or 
more). Therefore, we did not attempt to compare this method to our own.

\subsection{Sets of trajectories}

\begin{figure}
\resizebox{\hsize}{!}{\rotatebox{270}{\includegraphics*{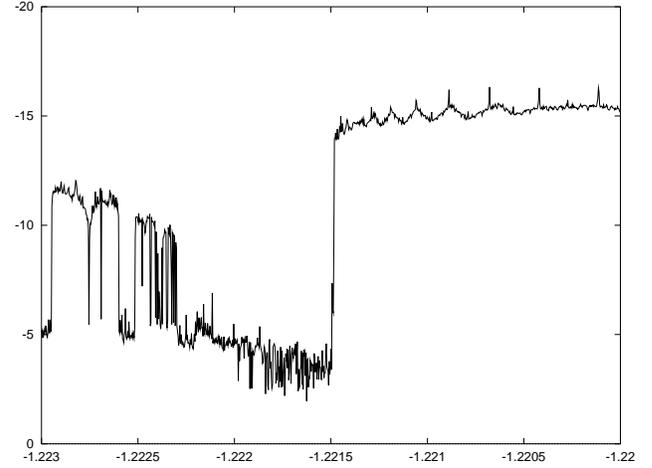}}}
\caption{The logarithm of the average amplitude of PSOD in a cross
section of 1000 trajectories near the 1/6 resonance in 
the standard map of eq.(6)} 
\label{helios}
\end{figure}

In order to circumvent the problem of calculating a trajectory for long times,
Contopoulos and Voglis (1997) proposed the use of the average value of $q$,
\begin{equation}
<q>=\frac{1}{N} \sum_{k=1}^N q_k
\end{equation}
If we keep N small, we can scan a wide area of initial conditions and map
its dynamical behavior. Trajectories in chaotic domains will have $<q>$ 
scattered around the value of the LCN of this domain, while trajectories 
in the ordered domain will have $<q>$ near zero. The method is very fast 
in distinguishing between ordered and chaotic domains. Fig.~9 of 
Contopoulos \& Voglis (1997) is a very good example of the results obtained 
by this method with only 10 iterations.

However, although the method is very good in scanning wide areas of
phase space for locating islands of order, it cannot give  reliable
results with so few iterations for a particular trajectory. In the case 
of stochastic trajectories, $<q>$ for $N=10$ varies so much, that it may 
yield a number as small as the one given for ordered trajectories. The 
situation is even worse in the case of
sticky trajectories (i.e. in the borders of islands). In order to 
decide on the character of such trajectories one needs considerably longer 
calculations.

Froeschl\'e \& Lega (1998) tested the FLI method along with the
method of twist angles (Contopoulos \& Voglis 1997), the 
frequency map analysis method (Laskar et al. 1992, Laskar
1993) and the sup-map method (Laskar 1990,
Froeschl\'e \& Lega 1996). Figs.~8a-d of their paper shows the results
of the four methods on a cross
section of 1\,000 trajectories near the hyperbolic point of the 1/6 resonance
of the standard map (Eq. 11) with $k=-1.3$. For the FLI method they
used 2\,000 iterations while for the other three methods 20\,000 iterations.
In order to produce unambiguous results, the trajectories in this test
are classified as ordered or chaotic by an appropriately selected
number/indicator. For the FLI method the authors have used as an indicator
the time necessary for the FLI to reach a value lower than $10^{-10}$.

In order to compare our method with these results we need also an one-number
indicator derived from the PSOD. As such we have selected here the
average value of $<P/P_{max}>$. The averaging is performed not over
all values but by ignoring the highest 1/6th and the lowest 1/6th of the 
amplitudes, the former being probably due to periodicities in sticky 
trajectories and the latter probably coming from numerical errors in the
integration and the calculation of the power spectrum.
Fig.~\ref{helios} shows the same cross section with Fig.~8a-d of
Froeschl\'e \& Lega (1998), using the above indicator taken after
8\,192 iterations. As we can see it gives essentially
the same information as the other four methods.
Note that the y-axis in Fig.~\ref{helios} is inverted for
easier comparison with Fig.~8a-d of Froeschl\'e \& Lega (1998).

\section{Application to asteroidal motion}

\begin{figure}
\resizebox{\hsize}{!}{\rotatebox{270}{\includegraphics*{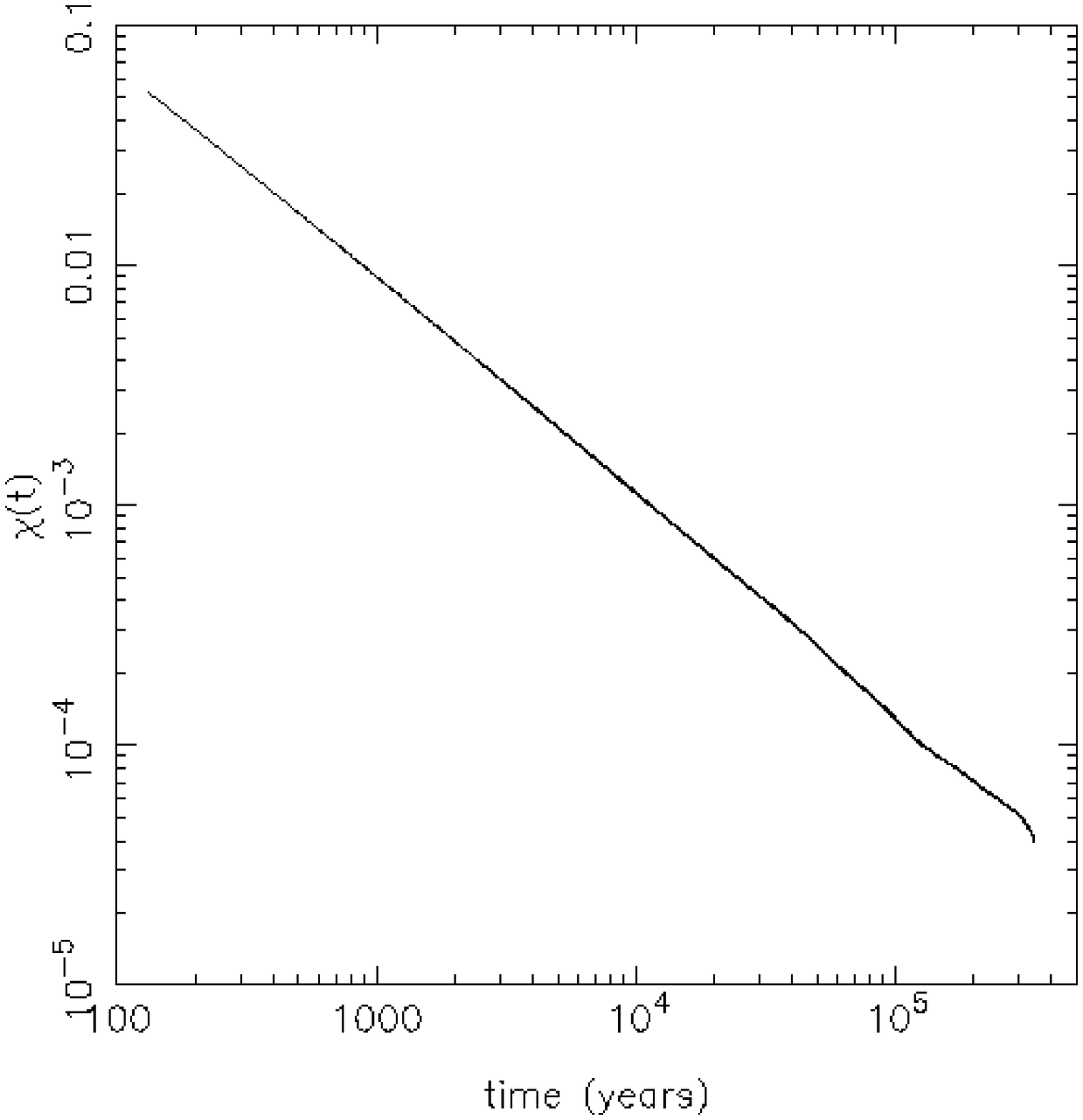}}
\hspace{1cm}          \rotatebox{270}{\includegraphics*{9135f18b.ps}}}
\vskip 0.1cm
\resizebox{\hsize}{!}{\rotatebox{270}{\includegraphics*{9135f18c.ps}}
\hspace{1cm}          \rotatebox{270}{\includegraphics*{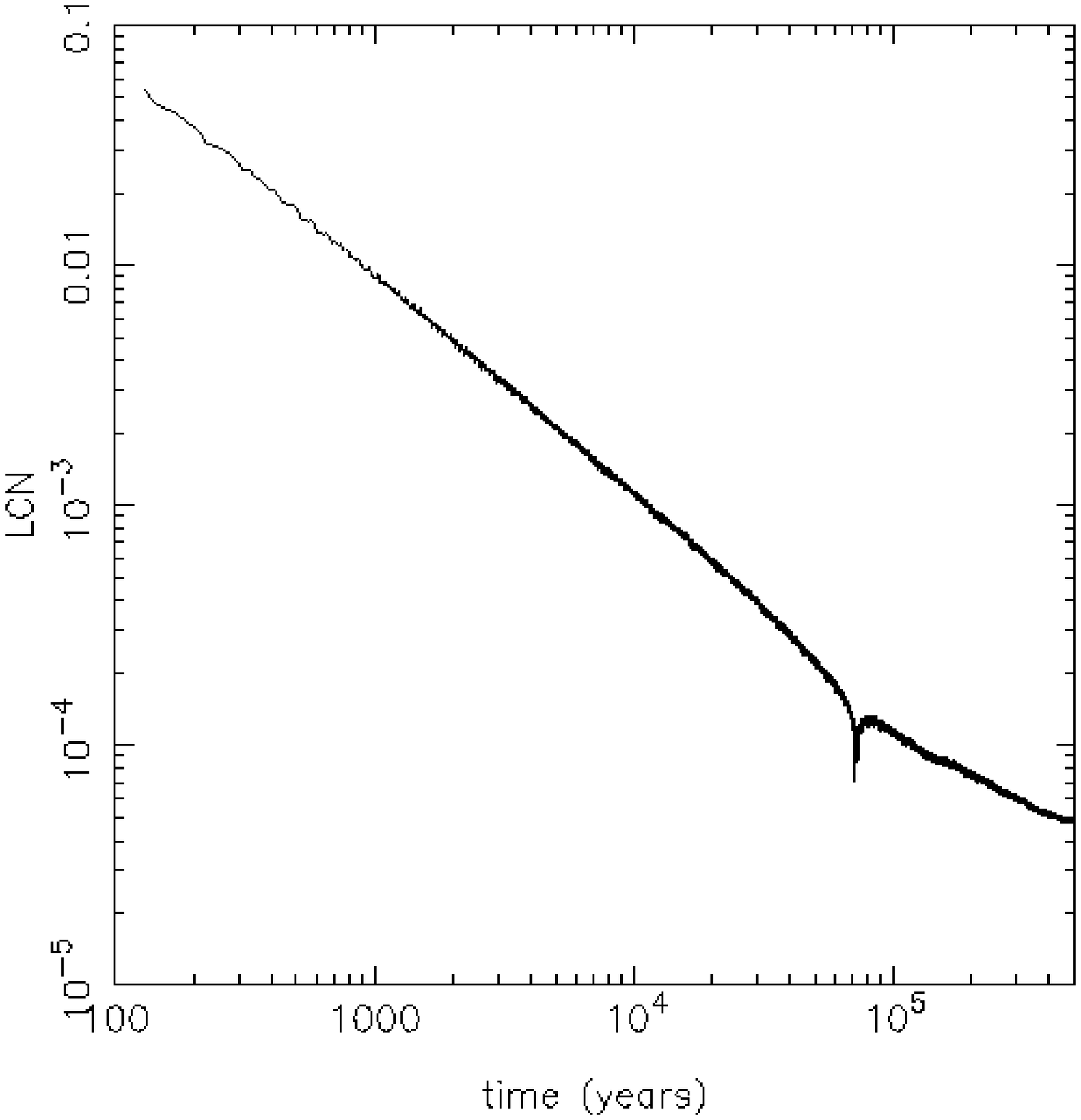}}}
\caption{The evolution of $\chi(t)$ of the trajectories of four
asteroids. Upper left, an 
ordered trajectory with initial $a=3$, $e=0$; Upper right a highly
stochastic
trajectory with $a=4$, $e=0.2$; Lower left, a stochastic trajectory with
stickiness and with initial
$a=3.64$, $e=0.08$; Lower right, another sticky orbit with initial
$a=3.63$, $e=0.08$} 
\label{3bp-lcn}
\end{figure}

\begin{figure}
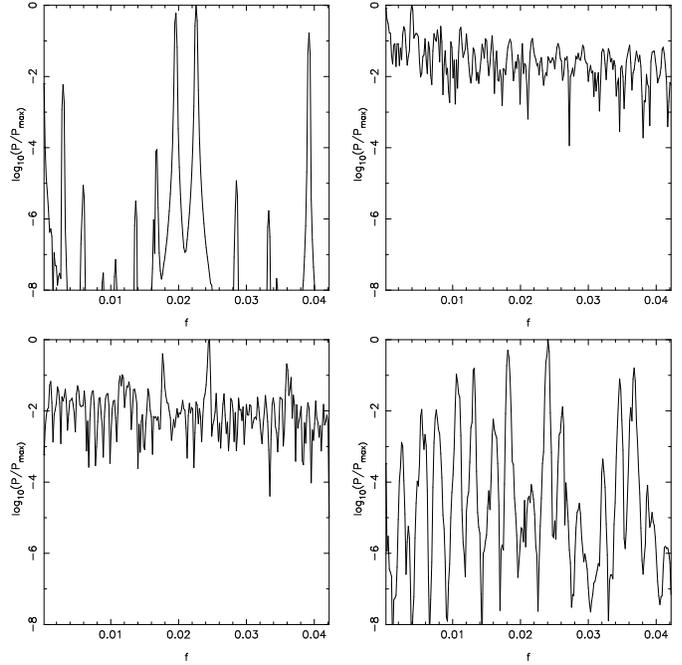

\resizebox{\hsize}{!}{\rotatebox{270}{\includegraphics*{9135f19a.ps}}
\hspace{1cm}          \rotatebox{270}{\includegraphics*{9135f19b.ps}}}
\vskip 0.1cm
\resizebox{\hsize}{!}{\rotatebox{270}{\includegraphics*{9135f19c.ps}}
\hspace{1cm}          \rotatebox{270}{\includegraphics*{9135f19d.ps}}}
\caption{The PSOD of the four asteroidal trajectories with N=512,
i.e. 6062 years} 
\label{3bp-psod}
\end{figure}

\begin{figure}
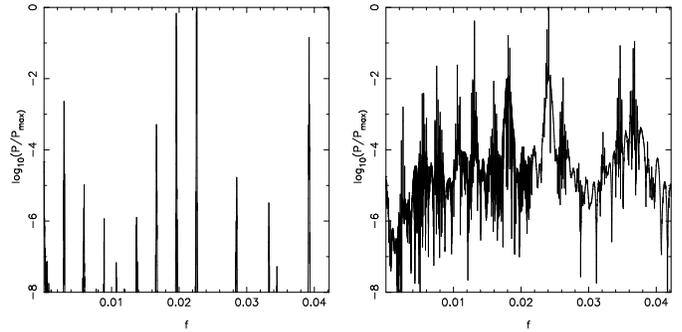

\resizebox{\hsize}{!}{\rotatebox{270}{\includegraphics*{9135f20a.ps}}
\hspace{1cm}          \rotatebox{270}{\includegraphics*{9135f20b.ps}}}
\caption{The PSOD, with N=8192, of the ordered asteroidal trajectory
(left) and the chaotic Helga clone with $a=3.63$, $e=0.08$ (right)} 
\label{3bp-psod2}
\end{figure}

As an application of our method to a problem of physical importance, we
shall use it in order 
to assess the chaotic or not nature of asteroidal trajectories. We use
a simplified model of the 
solar system,
namely the planar restricted three body problem where the
Sun and Jupiter move on elliptic trajectories around their center of
mass and an asteroid of
infinitesimal mass moves 
in the gravitational field of the two bodies. We calculate the PSOD
and the LCN using as 
renormalization time, $\Delta t$, the 
period of Jupiter, i.e. $\Delta t = T_J \simeq 11.86$ years.

Fig.~\ref{3bp-lcn} shows the $\chi (t)$ of the four trajectories
tested. The upper 
left frame belongs to an ordered trajectory starting with semi-major 
axis $a=3$ and
eccentricity $e=0$,
the upper right to a stochastic trajectory with initial elements $a=4$,
$e=0.2$. The other two frames correspond to trajectories with initial
$a=3.64$, $e=0.08$ (lower
left) and  $a=3.63$, $e=0.08$ (lower right) representing ``clones" of
the asteroid 
522 - Helga, which is a well-known example of ``stable chaos'' (Milani
\& Nobili 1992).

The PSOD of the four trajectories (N=512) is shown in Fig.~\ref{3bp-psod}.
As we can see, a few Jovian periods are sufficient to decide if the 
motion of a particular
asteroid is stochastic or ordered. In the case of the Helga clone
with $a=3.64$ (lower left of Fig.~\ref{3bp-psod})
the PSOD shows clearly a chaotic nature after only 512 Jovian periods
i.e. 6\,072 years, while even a rough calculation of the LCN
 needs at least $10^5$ years (see lower right frame of
Fig.~\ref{3bp-lcn}). The PSOD of the  $a=3.63$ Helga clone is rather
peculiar. Although it differs from that of the ordered orbit (upper
left), it is not clearly chaotic, unlike the PSOD of the other Helga 
clone (lower left). Nevertheless, if we take more points in our time series 
the chaotic nature of the orbit becomes apparent, as we can see in  
Fig.~\ref{3bp-psod2}. However, the spectrum can still be described as 
having a strong quasi-periodic component, something which is related to 
the peculiar dynamical nature of this orbit as discussed in Tsiganis et al. 
(2000b).

\section{Conclusions -- Discussion}

In the present paper we propose an alternative tool, which we call PSOD, 
for the characterization of the chaotic or not nature of
trajectories in conservative dynamical systems. The method is based 
on the frequency analysis of a time series, constructed by successive
records of the amplitude of the deviation vector of nearby 
trajectories.
As discussed in Section 2, such a ``mixed"\footnote{Since the method uses 
both the deviation vectors and frequency analysis it may be classified as
``mixed"} method 
is expected to have certain advantages. The reason is that the power 
spectrum of such a time series will contain all the characteristic 
frequencies of the motion in a properly ``weighted" ratio. 

The basic characteristic of the PSOD, seen in all three test cases (a 2-D 
mapping, a 2-D and a 3-D Hamiltonian system), is that  
\begin{itemize} 
\item for ordered trajectories the spetrum possesses only a few
high-amplitude peaks, the exact number of which depends not only on
the system but also on the particular orbit. Of course a small
amplitude noise level, due to the numerical procedure, is superimposed
on the spectrum, which diminishes as the length of the time series is
increased;
\item for chaotic trajectories the spectrum  has a  noisy pattern.
For weakly chaotic orbits a few high-amplitude peaks are also present.
Increasing the length of the time series, the spectrum tends to
a white noise spectrum which remains
practically unchanged for any (large \- enough) number of points (see Figs.
5 and 6).
\end{itemize}

Like most methods existing in the literature, it seems that the method 
performs better for maps than for flows. However, the results found for 
the three Hamiltonian flows tested (including the three-boody problem) 
show that the method can be applied to any system, no matter how many 
degrees of freedom are involved. We believe that the results may be 
significantly improved for flows, provided that proper analysis concerning 
the renormalization time is conducted. The difference between maps 
and flows is that, in the former case, isochronous records of dynamical 
quantities also mean studying the system on a surface of section. This is
not the case for flows and, thus, a more refined analysis on how to 
select a proper renormalization time has to be made.

We have shown that the sensitivity of the PSOD in testing single 
trajectories in maps is comparable to the FLI and the DSD methods.
This also makes the PSOD an efficient tool for scanning wide areas
of the phase space (see Section 4). For such purposes one would like
to have an one-number indicator for measuring chaos. The only uniquely 
defined measure of chaos is of course the LCN. Any other  
indicator should be in a one-to-one correspondence with the LCN, in 
order to give the same information. There is no guarantee that such an  
indicator can be based on the frequency content of the PSOD, as 
chaotic orbits with similar LCNs may have a different frequency distribution.
Further analysis of the properties of the PSOD has to be performed in
order to decide whether such an indicator can be found.

\begin{acknowledgements}
The authors would like to acknowledge the constructive comments of the
anonymous referee.
\end{acknowledgements}

\end{document}